# Self-organization of dissipative and coherent vortex structures in non-equilibrium magnetized two-dimensional plasmas

O Bystrenko and T Bystrenko

Bogolyubov Institute for theoretical physics, Kiev, Ukraine

e-mail: obystr@bitp.kiev.ua

The properties of non-equilibrium magnetized plasmas confined in planar geometry are studied on the basis of the first principle microscopic Langevin dynamics computer simulations. The non-equilibrium state of plasmas is maintained due to the recombination and generation of charges. The intrinsic microscopic structure of non-equilibrium steady-state magnetized plasmas, in particular, the inter-particle correlations and self-organization of vortex structures are examined. The simulations have been performed for a wide range of parameters including strong plasma coupling, high charge recombination and generation rates, and intense magnetic field. As is shown in simulations, the non-equilibrium recombination and generation processes trigger the formation of ordered dissipative or coherent drift vortex states in 2D plasmas with distinctly spatially separated components, which are far from thermal equilibrium. This is evident from the unusual properties of binary distributions and behavior of the Coulomb energy of the system, which turn out to be quite different from the ones typical for the equilibrium state of plasmas under the same conditions.

**Introduction.**

Cooperative phenomena in non-equilibrium open systems, beginning with the works by Belousov and Zhabotinsky [1, 2], attracted permanent considerable interest of researchers. The general theoretical basis for the description of ordering in such systems has been given by Prigogine [3,4] and Haken [5]. The theory of self-organization in open systems proposed by Haken, the synergetics, suggests the consistent probabilistic description of such systems based on Fokker-Planck equations, which in turn originate from the stochastic Langevin dynamics. However, in actual practice, the theoretical models for ordered dissipative structures (DS) occurring in particular open systems incorporate, as a rule, additional phenomenological assumptions or approximations (see, for instance, [6]). Evidently, this fact is associated with the complexity of real systems, which makes the consistent accurate implementation of the theory rather difficult. In view of this, the outstanding challenge in this field would be to describe cooperative phenomena in more or less simple specific systems on a basis of the Langevin particle dynamics by means of computer simulations. In particular, this would provide possibility to verify the validity of the basic principles of synergetics in direct numerical simulations.

In this work we study the properties of vortex formation in two-dimensional (2D) magnetized plasmas, where the non-equilibrium is maintained due to the processes of recombination and generation of charges, by means of computer simulations based on first principle microscopic particle dynamics. Vortex structures occurring in magnetized plasmas attracted considerable attention of researchers during decades. The vast majority of theoretical works in this area is based on the continuous models, such as the classical Hasegava-Mima theory [7] or the drift-Poisson model [8]. Due to the computer simulations based on these models, which are performed predominantly for the case of cylindrical geometry, a significant progress has been achieved, in particular, in the description of the evolution of vortex structures from the initial turbulent state, and the explanation of formation of vortex crystals and other experiments with non-neutral magnetized electron plasmas in the electron columns [9-11 ].

The above mentioned continuous models originate in the end from the plasma hydrodynamics based on the mean field concept. As the result, one of the restrictions for them is that the plasma must be ideal (i.e. weakly coupled). This means that the continuous approaches completely neglect the effects associated with inter-particle correlations in plasmas, and provide no way to study intrinsic microscopic plasma structure, particularly, in the case when the plasma coupling is not weak. At the same time, as has been demonstrated by recent experimental and theoretical studies of strongly coupled plasmas, these latter may differ drastically in properties from the weakly coupled plasmas [12]. The microscopic computer simulations based on first principle particle dynamics are not subject to the above drawback and provide tools to examine the correlations and structure of vortices in magnetized plasmas. As far as we know, such simulations up to this time have yet not been performed.

As the relevant physical system associated with this problem, we assume to be the planar heterostructures in semiconductors like quantum wells, where the electron-hole plasma is confined in planar 2D geometry. The properties of such systems are intensively investigated in recent years from both experimental and theoretical points of view [13, 14]. The crystal lattice responsible for the dissipation due to the scattering of charge carriers represents in this case a natural heat bath, the presence of which is taken into account in Langevin equations.

Since the external magnetic field does not contribute to the Gibbs' distribution for a system of point charges, the equilibrium properties (thermodynamic averages) of such a system do not depend on its presence. The presence of magnetic field manifests itself only in the non-equilibrium processes, in particular, in the kinetic properties. Note, that the microscopic description of two-component plasmas (TCP) suggests a natural source of non-equilibrium, namely, the recombination. The need to take into consideration the processes of recombination is connected with the fact that the TCP consisting of point charges can not be in thermal equilibrium because of the classical collapse of oppositely charged components [15]. By the way, this is its one more distinctive feature in contrast to continuous models. For this reason our microscopic model includes, along with the point plasma charges interacting via Coulomb forces and the transversal magnetic field, the processes of charge recombination (and generation, if steady states have to be examined).

Let us formulate now the main distinctive features which distinguish our approach from the previous works in this field. Firstly, we employ the microscopic description based on Langevin dynamics allowing for the dissipation, which is applicable in the case of strong correlations. Secondly, we take into account the recombination and generation of charges which enables one to examine the associated non-equilibrium processes. And, finally, we investigate the case of two-dimensional (2D) planar geometry rather than cylindrical one.

**1. Formulation of the dynamic equations.**

Thus, we consider the TCP consisting of point charges interacting via Coulomb interaction. It is confined in 2D planar (in *XOY* plane) geometry and exposed to external constant magnetic field oriented perpendicularly, along *Z*-axis. The equations of motion for the particles we employ are the Langevin dynamics (LD) equations [16]

$$m_i \frac{d^2 \vec{r}_i}{dt^2} = \vec{K}_i - \gamma_i v_i + \vec{F}_i(t) \qquad (1.1)$$

Here $\vec{r}_i$ and $m_i$ is the radius-vector and the mass of the *i*-th particle respectively, $\vec{v}_i \equiv \frac{d\vec{r}_i}{dt}$ is its velocity, and

$$\vec{K}_i = -\nabla_i U_{Coul} + \vec{K}_{i,ext} \qquad (1.2)$$

is the regular force acting on the *i*-th particle. The quantity $U_{Coul} = \frac{1}{2} \sum_{i \neq j} \frac{z_i z_j e^2}{|\vec{r}_i - \vec{r}_j|}$ is the Coulomb energy of the system of particles (where *e* is the absolute value of the electron charge, $z_i = \pm 1$ is the charge number of *i*-th particle), and $\vec{F}_i$ is the stochastic Langevin force (acting on *i*-th particle) due to the presence of heat bath defined by the correlator

$$\left\langle F_{i,\alpha}(t) F_{j,\beta}(t') \right\rangle = 2\gamma_i k_B T \delta_{ij} \delta_{\alpha\beta} \delta(t-t') \qquad (1.3)$$

with $\gamma_i$ being the relevant friction coefficient. Thus, we assume that the heat bath is in the thermal equilibrium at the temperature $T$. Subscripts $\alpha, \beta \equiv x, y, z$ denote the Cartesian components of corresponding vectors, and $k_B$ is the Boltzmann constant. In our case the external force $\vec{K}_{i,ext}$ includes, along with possible external potential fields, the Lorentz force $\vec{K}_{i,L} = \dfrac{ez_i}{c}\left[\vec{v}_i \times \vec{B}\right]$, where $\vec{B}$ is the magnetic induction; c is light velocity.

Since in this work we study the self-organization processes at long time scales, $t \gg m/\gamma$, we may simplify the LD equations by setting the inertial term equal to zero. After simple transformations, Eq.(1.1) may be reduced to the overdamped form

$$\vec{v}_i = \frac{1}{\gamma_i}\left\{\vec{K}_i + \vec{F}_i(t)\right\} \tag{1.4}$$

We see that the overall velocity naturally splits in two parts, the drift $\vec{v}_{i,drift} = \dfrac{1}{\gamma_i}\vec{K}_i$ associated with the regular part of the forces, and diffusion $\vec{v}_{i,diff} = \dfrac{1}{\gamma_i}\vec{F}_i(t)$ associated with stochastic forces.

The equations of motion (1.4), which we shall hereafter refer to as Brownian dynamics (BD), determine the particle dynamics in the drift-diffusion limit. By explicitly taking into account the magnetic field $\vec{B} \parallel OZ$ one can reformulate the BD equations into the form

$$\vec{v}_i = \frac{1}{\gamma_i(1+\beta_i^2)}\left\{\vec{P}_i + \beta_i\left[\vec{P}_i \times \vec{e}_z\right]\right\} \tag{1.5}$$

where $\vec{e}_z$ is the unit vector giving the orientation of the OZ-axis. We introduced here the dimensionless parameter $\beta_i = \dfrac{ez_i B}{c\gamma_i}$ specifying the field strength in relation to the friction, and the total force acting on i-th particle

$$\vec{P}_i = -\nabla_i U + \vec{F}_i(t) \tag{1.6}$$

where the potential energy $U = U_{Coul} + U_{ext}$ of many-particle system includes, along with Coulomb energy, the energy in possible external potential fields. The above equations (1.5) determine the overdamped molecular dynamics (MD) of charged particles in a thermostat in the presence of magnetic field. Exactly these equations (for the case where the 2D particle dynamics is confined within the XOY plane) were employed predominantly in our computer simulations in order to investigate specific problems under consideration. Note that the equations (1.5) are quite equivalent to the ones obtained in the paper [17] dealing with closely related problems.

It is worth mentioning that from Eq.(1.5) one can easily obtain the mean-square displacement for a charged particle moving in the presence of magnetic field $\vec{B} \parallel OZ$ on 2D XOY-plane in thermostat,

$$\langle r^2 \rangle = \frac{4Dt}{(1+\beta^2)} \tag{1.7}$$

recovering the known result that the diffusivity in this case reduces by the factor $(1+\beta^2)$ [18]. To establish the correspondence between friction $\gamma$ and diffusivity D, we used herewith the Einstein relation

$$\gamma D = k_B T. \tag{1.8}$$

Let us say now a few words about continuous probabilistic description of the problem under consideration associated with the BD particle dynamics. To obtain the relevant drift-diffusion equations for the positive and negative plasma densities $n_{p,n}$, one can use the method given in the work [16]. Explicit allowance for the presence of magnetic field yields

$$\frac{\partial n_{p,n}}{\partial t} = -\frac{1}{\gamma_{p,n}(1+\beta_{p,n}^2)} \operatorname{div}\left\{n_{p,n}\left(\vec{K}_{p,n} + \beta_{p,n}[\vec{K}_{p,n} \times \vec{e}_z]\right)\right\} + \frac{D_{p,n}}{(1+\beta_{p,n}^2)}\Delta n_{p,n}, \tag{1.9}$$

where $\vec{K}_{p,n} = -ez_{p,n}\nabla\phi$, and for the potential $\phi$ one needs to solve the associated Poisson problem

$$\Delta\phi = -4\pi e(z_p n_p + z_n n_n). \tag{1.10}$$

As we see, the essential difference between the continuous drift-diffusion and microscopic BD approaches is that the former is based on the mean-field concept for the potential $\phi$ and provides no possibility to study the inter-particle correlations.

Note that, as one would expect, in the absence of magnetic field ($\vec{K} \neq 0, \beta = 0$) Eq. (1.9) reduces to the conventional Smolukhovsky equation (i.e., Fokker-Planck equation in the coordinate space) [16], and in the absence of the Coulomb forces but in the presence of magnetic field ($\vec{K} = 0, \beta \neq 0$) we obtain the diffusion equation with the modified diffusivity, i.e., the result consistent with Eq.(1.7) for a single charged particle. At the same time, in the limit $\beta \to \infty$ one can easily obtain from Eq.(1.9) the drift-Poisson model mentioned above. We stress these points, since they evidence for the correctness of the basic dynamic equations (1.5) used in our computer simulations.

## 2. Description of numerical simulations.

Note that in actual simulations we used dimensionless form of equations obtained by employing the physical units relevant to the problem and dimensionless parameters. In what follows, we use size $L$ of the simulation square on a 2D plane as a length unit, and $t_0 = L^2/D_p$ (typical diffusion time over the distance $L$) as a time unit. Respectively, associated dimensionless variables are: length $\vec{\rho} \equiv \vec{r}/L$ and time $\tau \equiv t/t_0$. The other dimensionless parameters relevant to the problem, which were used in simulations are as follows. The field strength parameters mentioned in the preceding Section,

$$\beta_{p,n} \equiv \frac{ez_{p,n}B}{c\gamma_{p,n}} \tag{2.1}$$

for positive and negative plasma components ($z_{p,n} = \pm 1$). The coupling parameter,

$$\chi \equiv \frac{e^2}{k_B T L} \tag{2.2}$$

which specifies the potential-to-kinetic energy ratio in the system and relates in 2D case to the commonly used in physics of strongly coupled plasmas coupling constant as $\Gamma \approx \sqrt{2\pi N}\chi$ (Here $N = N_p = N_n$ is the number of particles in the simulation square for plasma components). Also, we used in simulations the dimensionless recombination coefficient, $\sigma \equiv c_r/D_p$ with $c_r$ being the given 2D recombination coefficient (note, that in 2D case the recombination and diffusivity are of the same dimensionality), and the ratio of diffusivities $\theta = D_n/D_p$.

Thus, in numerical simulations BD equations (1.5) were solved for overall neutral systems consisting of equal numbers of oppositely charged point particles, within a simulation square of the unit size, with periodically repeating in *X*- and *Y*-directions image configurations. To allow for the periodic boundary conditions, the nearest image approximation has been used, which is usual for particle MD or Monte Carlo simulations [19]. The pair Coulomb interaction between particles has been slightly modified in simulations at short distances (for $\rho \leq \rho_c$),

$$\frac{V_{Coul}(\rho_{ik})}{k_B T} = \chi \frac{z_i z_k}{\rho_{ik}} erf(\rho_{ik}/\rho_c) \tag{2.3}$$

in order to eliminate the singularity. In simulations we choose $\rho_c$ to be rather small, at least by order less than the typical radius of recombination (see below).

As concerns the description of recombination, the main requirement is to reproduce with good accuracy the overall recombination process in accordance with the well-known relation

$$\frac{dN_{p,n}}{d\tau} = -\sigma N_p N_n. \tag{2.4}$$

For this purpose, we specify the probability for a pair of opposite charges placed at a distance $\rho$ to recombine during the time span $\Delta\tau$ as

$$W(\rho, \Delta\tau) = \exp\{-(\rho/\rho_0)^2\} \tag{2.5}$$

Here we introduced the recombination radius $\rho_0 = \sqrt{2\bar{D}\sigma\Delta\tau}$, where $\bar{D} = 1/(1+\beta_p^2) + \theta/(1+\beta_n^2)$ has the meaning of the modified diffusivity. The expression for the probability (2.5) can be obtained as an approximate solution of the relevant recombination-diffusion equation for a single pair of particles for short time intervals $\Delta\tau$. As was tested in computer simulations, this approach allows one to describe with satisfactory accuracy the recombination process given by Eq.(2.4).

Along with recombination, we took into consideration the process of particle generation, for instance, induced by UV radiation or as a result of charge injection. In all the simulations, we assumed the relevant source intensity $I$ to be uniformly distributed over the simulation square, without spatial correlations, and constant with time. It was supposed therewith that the particles are created randomly in time and space and in pairs, so that the overall charge neutrality is conserved.

The random Langevin forces were generated in simulations as to satisfy the relations (1.3), in a manner given in Ref.[20]. The amplitude of the random force $\vec{F}(t)$ acting on a particle during the time span $\Delta t$ is specified in 2D case by the Gaussian probability distribution

$$w(\vec{M}) = \frac{1}{4\pi\gamma^2 D\Delta t} \exp\left(-\frac{M^2}{4\gamma^2 D\Delta t}\right) \tag{2.6}$$

for the momentum $\vec{M}$

$$\vec{M}(\Delta t) = \int_t^{t+\Delta t} \vec{F}(t)dt \tag{2.7}$$

to be transferred to this particle. We omitted here the obvious subscripts for simplicity.

The general method of simulations was as follows. The overall time interval of a simulation run is divided into sufficiently short time steps ($\Delta\tau = 10^{-1}\ldots 10^{-4}$). Beginning with initial given particle configuration and set of parameters, Eqs.(1.5) are solved on each step with the current number of particles and current (generated for this time step) amplitudes of random Langevin forces. At the end of the time step, the probabilities for each pair of particles in the configuration to recombine, and for new pairs of particles to be created are evaluated. Then, according to theses, the changes into the configuration are introduced. Therefore, the total number of particles, and, respectively, the number of equations of motions (1.5) changes with time. After that, the procedure repeats until the final time point is achieved. Note, that the cooperative phenomena, which we are going to examine, manifest themselves at rather long time intervals, on the order of $\tau \cong 10^2 \ldots 10^3$. For this reason one should expect that the results of simulations should be not sensitive to the choice of time step, which is an arbitrary parameter of simulations. In actual fact, the magnitude of a time step is determined by the requirement that the displacement of each particle must be much less than the typical spatial irregularity of fields. Accordingly, higher magnitudes of magnetic field admit longer time steps (as evidenced by Eq.(1.7)), whereas the strong coupling makes necessary the shorter time steps. The configurations obtained in the above way, are written then successively into a file for subsequent analysis.

In simulations there were monitored the total number of particles, the Coulomb energy per particle, binary distribution functions. Also, we evaluated the kinetic energy of drift motion per particle for both components, which we define as

$$E_{p,n}^{kin}(t) \equiv \frac{1}{2}\left\langle\vec{v}_{p,n}^{\,2}(t)\right\rangle, \tag{2.8}$$

and the spectral distribution of the above kinetic energy over a given time interval defined as

$$S_{p,n}(\omega) = A_{p,n}\left\langle\vec{v}_{p,n}^{\,2}(\omega)\right\rangle, \tag{2.9}$$

where $\vec{v}_{p,n}(\omega)$ are the Fourier amplitudes for velocities of drift motion over that time interval, $A_{p,n} = 1/\sum_\omega \left\langle\vec{v}_{p,n}^{\,2}(\omega)\right\rangle$ is the relevant normalization coefficient. The brackets denote as usual the averaging over the ensemble of particles. The spectral density $S$ enables one to estimate the extent of coherency of rotational motion of particles.

It should be mentioned that, in order to check the accuracy of simulations, we performed a large number of tests. Most important of them, which have the physical meaning, are the drift and diffusion tests, recombination tests mentioned above, and comparative studies based on LD and BD simulations. In drift and diffusion tests we checked the consistency of the results for average drift and mean-square displacement of a single charged particle in magnetic field obtained within LD, BD and genuine microscopic MD, where the presence of the heat bath was simulated by random scattering of particles. In particular, the validity of Eq.(1.7) in the drift-diffusion limit has been verified. In comparative LD and BD tests the agreement between the results obtained within these approaches for many particle TCP in magnetic field has been checked.

In addition, we performed a number of tests of the properties of equilibrium one-component plasmas (OCP). As is known, OCP are capable of forming Coulomb liquids or crystal lattice at strong coupling. The usual methods of computer simulations for these are Monte Carlo or MD simulations (see, for instance, [21]). The BD simulations demonstrated the possibility to reproduce these properties (we checked the formation of a 2D hexagonal lattice, binary distributions, etc.). This means that the strong correlations are correctly described within BD approach.

## 3. Numerical simulations. Discussion of the results.

BD simulations for the above formulated system have been performed with the total number of particles $N_{tot} \cong 100\ldots 2000$ for the following range of parameters. Namely, field strength parameter $\beta = 10\ldots 500$; ratio of diffusivities $\theta = 0;1;10$; coupling parameter $\chi = 0.01\ldots 20$ (which corresponds approximately to the range of the plasma coupling $\Gamma \cong 0.1\ldots 500$); recombination $\sigma \cong 1\ldots 10^3$; source intensity $I = 0\ldots 1000$. The most important from theoretical point of view results have been obtained for rather extreme parameters as regards their relevance to the real physical planar semiconductor systems. Below we will discuss this issue in more detail, but first we present the results of a number of runs with most significant typical features.

The first series of runs has been performed for the following parameters, $I = 200$; $\sigma = 200$; $\beta_p = 500$; $\theta = 10$; $\chi = 2 \cdot 10^{-2}$. The simulations started from the initial particle configuration with $N_p = N_n = 510$ and random uniform spatial distribution. The results are given in figures. In Figs.1 and 2, a series of subsequent snapshots of plasma configurations is given illustrating the formation of a vortex structure. The associated behavior of the particle number and the Coulomb energy per particle is given in Figs.3 and 4. The formation of a steady state occurs during the time span of the order $\tau \approx 100\ldots 500$ as the result of building up fluctuations accompanied by creation of small vortices, which then merge together. The examination of the configuration in Fig.2 reveals that the opposite plasma components in the final steady state are to considerable extent spatially separated. The behavior of spectral density for the drift kinetic energy manifests the peaks, which correspond to the rotation frequency of the majority of plasma particles (Fig.5). Note that the more mobile negative component (with higher diffusivity) rotates at higher speed and forms a more compact and pronounced vortex, than the slow positive component. This is also evident from the behavior of the kinetic energy of drift motion in Fig.6, where the acceleration of plasma components in the process of self-organization is observed. In Fig.7, we give the behavior of binary distribution functions, which show the tendency of like-charged particles to form compact clusters, whereas the oppositely charged particles tend to be spatially separated. Remarkably, these distributions indicate the high probability of formation of like-charged pairs of particles at close separations, which, in actual fact, turn out to be micro-vortices. Note, that this behavior of pair distributions is quite opposite to that typical for the equilibrium TCP. The unusual behavior of the Coulomb energy should be as well pointed out. Typically, the Coulomb energy per particle in equilibrium TCP is approximately equal to $\cong -\Gamma$, while for the vortex structures observed in these simulations it is positive and considerably exceeds the coupling constant (for the final steady state its average value was estimated as $\Gamma \approx 1.5$). Obviously, all the above mentioned evidences, that the plasma system is far from equilibrium, which relates in the end to the spatial self-organization (separation of components). Note that this plasma state can be identified as rather non-ideal, since for the ideal plasma must be $\Gamma \ll 1$.

Similar structures have been observed in numerical experiments at different plasma parameters. The most essential requirement for the above vortex self-organization is that the recombination must be rather intensive ($\sigma \gg 1$), and the magnetic field strong ($\beta \gg 1$). We regard the vortex structures obtained in these runs as DS, since they appear as the result of a process of self-organization in open dissipative systems, where the non-equilibrium is maintained due to the recombination and generation of particles, and exhibit marked signs of ordering. Note, that we can regard the pair distributions or spectral densities for velocities as the order parameters for these DS. One of the most distinguishing signs of vortex growth is the above mentioned unusual behavior of the Coulomb energy of the system.

In the next series of runs we examined the relaxation of DS obtained in previous runs. We call here the relaxation a free decay of an initial plasma state in the absence of sources of particle generation (i.e. with no pumping). Naturally, this non-equilibrium process is induced, in particular, by plasma recombination and dissipation. The simulations started from the final configurations of a steady-state DS obtained a manner described above, with the same parameters, except for the source intensity, $I = 0$. The results presented in Figs. 8-13 indicate that, as the result of relaxation, the magnetized TCP forms an almost steady collective vortex state with very distinctly spatially separated plasma components and high extent of coherency. The snapshot of configuration for this state is given in Fig. 8. As is seen from the behavior of the particle number and Coulomb energy (Figs. 9 and 10), during the short time span $\Delta \tau = 300...310$, the processes of intensive recombination are dominating accompanied by the increase in the Coulomb energy. After that, the system acquires a state of relatively high coherency, which is evidenced, in particular, by the direct visual observations of particle configurations. Then, the system slowly loses its kinetic energy due to dissipation (Fig. 11). In particular, the retardation manifests itself in lowering the rotation frequencies of plasma components (as compared to the above case of steady-state DS). However, the extent of coherency of rotational mode increases therewith, and the kinetic energy tends to concentrate in a narrow frequency band of the spectral distributions (Fig. 12). At this stage the recombination becomes rather insignificant, obviously, due to the spatial separation of plasma components. We performed a number of test runs starting from this coherent vortex state, after making the recombination negligibly small ($\sigma \simeq 10^{-1}; 10^{-3}; 10^{-5}$).

These runs have shown that the behavior of the coherent vortex mode does not essentially depend on the intensity of recombination. The radial distributions exhibit qualitatively same features, as for the DS (Fig.13). In general, the properties of vortex DS and free coherent vortex mode in 2D magnetized plasmas turned out to be similar. The main difference is that the plasma rotation in DS is much less coherent.

Thus, we observed in simulations the formation of DS in the open system, namely, the magnetized 2D TCP. The essence of the phenomenon is that due non-equilibrium recombination and generation processes, predominantly the drift vortex (rotational) collective mode is excited. Apparently, this steady state of DS can exist as long as the pumping and the recombination are present in the system. However, after being excited, the above collective mode can exist rather long by itself even in the case that the sources of non-equilibrium (pumping and recombination) are removed. The life time of this free mode is determined, evidently, by the dissipation present in the system. In this connection, we find it appropriate to draw the important analogy with the classic example of an open DS, the laser. In a laser, as well, there exists a collective mode represented by a standing electromagnetic wave in a mirror resonator. Once being excited, it decays due to the dissipation, and in the absence of dissipation and losses (e.g., via mirrors) would exist infinitely long. However, in order to trigger the process of self-organization, i.e., to excite this mode, the pumping and dissipation in an active media is needed. In our case the role of the dominating collective mode plays the vortex rotational mode, and the generation and recombination are responsible for its excitation.

Let us say a few words about the possibility of experimental observation of the vortex DS predicted in our computer simulations in planar semiconductor structures. We performed simple estimates for more or less plausible for semiconductors and the problem as a whole range of parameters, namely, square size $L \approx 1...3 \mu m$; surface charge density $n \approx 10^{10} cm^{-2}$; electron/hole diffusivity $D \cong 10^2...10^3 \ cm^2 / \sec$; temperature $kT \cong 0.001...0.05 \ eV$; magnetic field $B \cong 10...20 \ Tl$; source intensity (pumping) $I \cong 10^{18}...10^{19} \ cm^{-2} \sec^{-1}$; typical time scale $t \cong 10^{-9}...10^{-10} \sec$. They have shown that the most of parameters, though being rather extreme, still can be achieved by employing strong magnetic fields and the semiconductors with high carrier mobility (diffusivity) like GaAs. The problem may arise with the 2D recombination coefficient, which in real semiconductors seems to be much less than that needed to initiate the process of vortex formation. However, we expect that the implementation of special experimental processing like doping with recombination-enhancing traps-impurities could remedy this issue.

**Conclusions**

To conclude, we study the formation of vortex structures in non-equilibrium magnetized plasmas, with allowance for the charge recombination and generation, confined in a planar geometry, on the basis of microscopic first principle Langevin dynamics computer simulations. In simulations, there have been examined the behavior of the Coulomb energy, kinetic energy of drift motion and its spectral distributions, and the intrinsic microscopic structure of vortices. The simulations have shown that, for sufficiently strong magnetic fields and intensive recombination and pumping, the magnetized 2D plasma forms steady-state vortex dissipative structures maintained due to non-equilibrium recombination and generation processes. The plasma state in these DS is far from equilibrium, which is evident from the properties of binary distributions and behavior of the Coulomb energy of the system. These latter are distinctly different from the ones typical for equilibrium plasmas and indicate pronounced spatial separation of positive and negative plasma components. The investigation of the relaxation of vortex dissipative structures in the absence of pumping reveals the formation of slowly decaying rotational vortex mode with the high extent of coherency. Due to the spatial separation of components, the recombination in this coherent state is negligibly small. We expect that the predicted dissipative and coherent vortex structures could be observed in such physical systems as the electron-hole plasmas in semiconductor hetero-structures, e.g. quantum wells, or other planar plasma systems with dissipation.

FIGURES

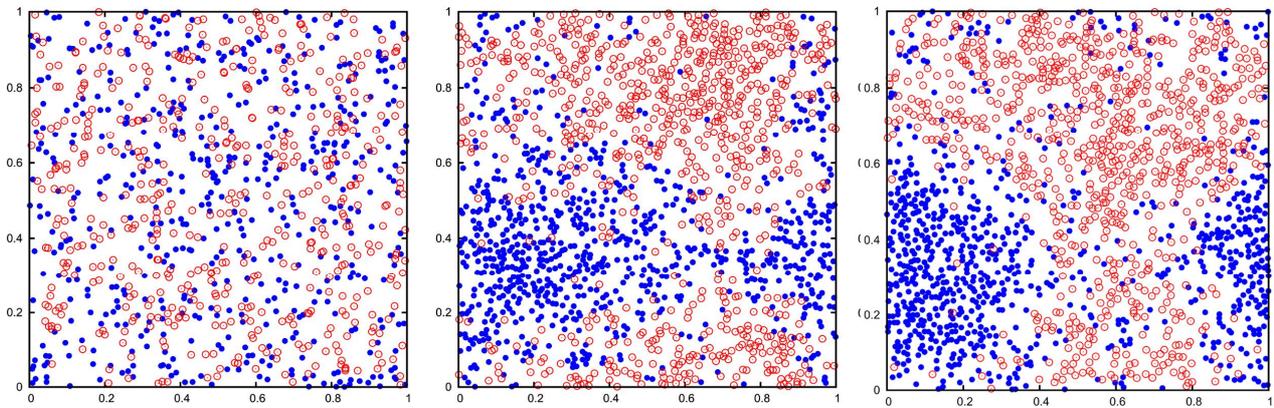

Fig.1. Succesive plasma configurations for the time points $\tau = 0; 100; 200$ (from the left to the right). Open circles in red denote the positive plasma component, and filled circles in blue denote the negative plasma component. The process of spatial separation of plasma components is accompanied by the growth in rotation of plasma clusters.

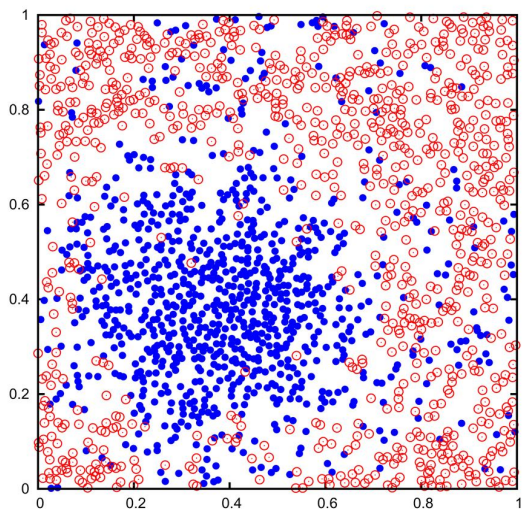

Fig.2. Steady-state vortex DS formed at $\tau = 300$ as the result of the evolution given in Fig.1.

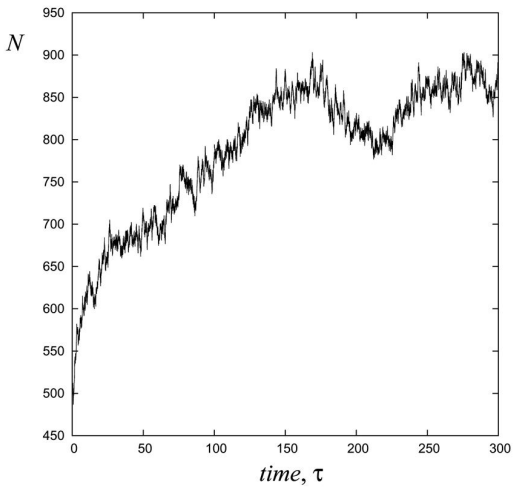

Fig.3. Number of particles in simulation square as a function of time.

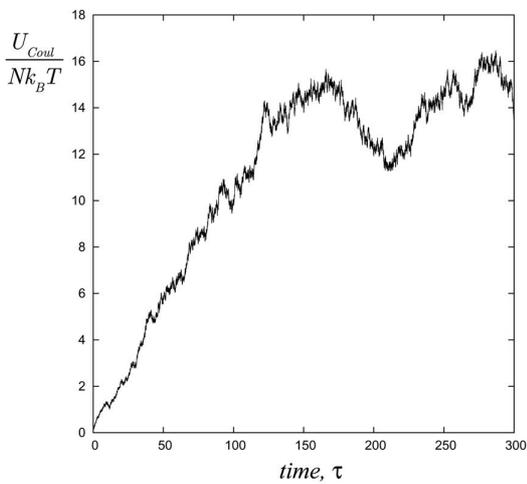

Fig.4. Behavior of the Coulomb energy of a 2D plasma system in the process of the formation of vortex DS.

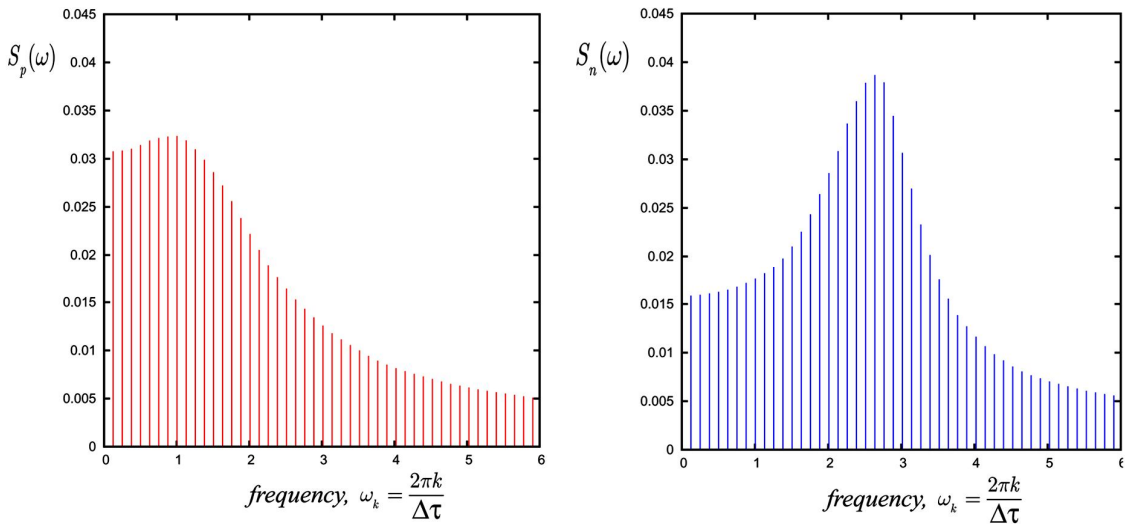

Fig.5. Spectral distributions for kinetic energy of drift motion for positive (left) and negative (right) plasma components evaluated for the steady vortex DS over the time interval $\Delta\tau = 250\ldots300$.
time.

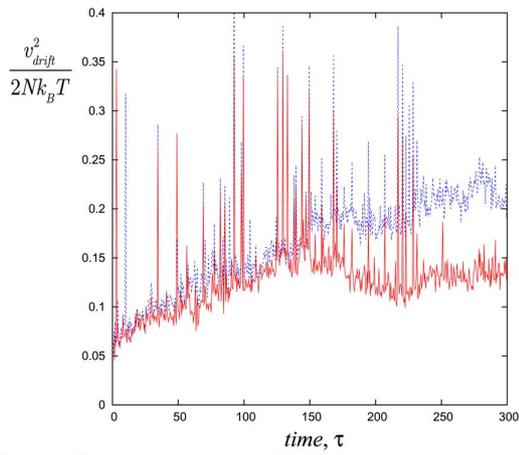

Fig.6. Kinetic energy of drift motion for positive (solid in red) and negative (dashed in blue) plasma components during the process of self-organization of vortex DS. The sharp peaks occur due to the Coulomb collisions at close inter-particle distances.

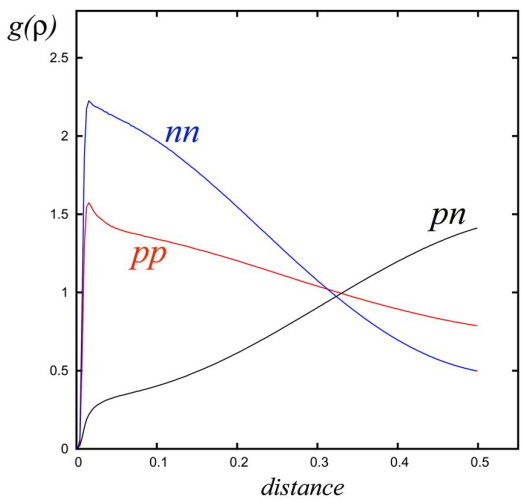

Fig.7. Binary radial 2D distributions for positive-positive (pp, in red), negative-negative (nn, in blue), and positive-negative (pn, in black) plasma components averaged over the time span $\tau = 250\ldots 300$.

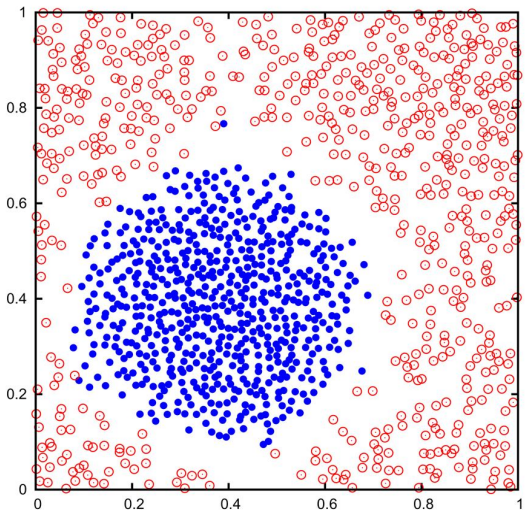

Fig.8. Coherent vortex state formed at $\tau = 400$ after the relaxation period. The pumping has been turned off at $\tau = 300$.

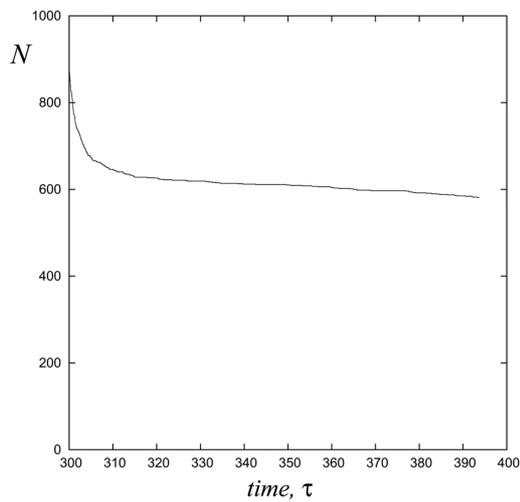

Fig.9. Number of particles as a function of time during the process of relaxation.

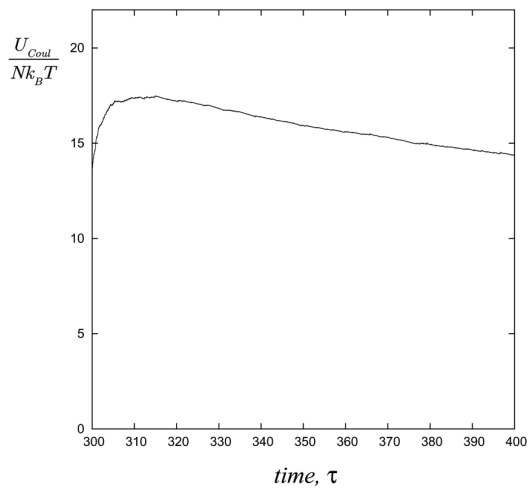

Fig.10. Behavior of the Coulomb energy of the system during the relaxation. The pumping has been turned off at $\tau = 300$.

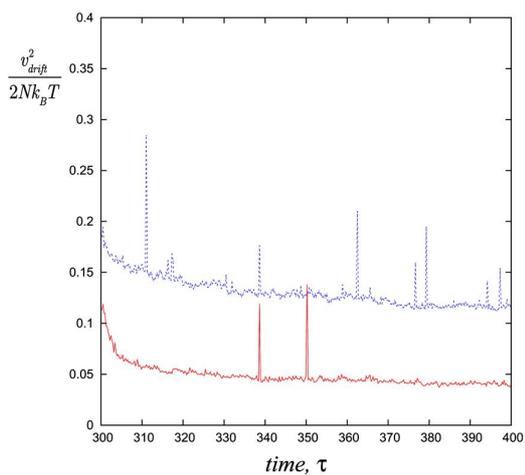

Fig.11. Kinetic energy of drift motion for positive (solid in red) and negative (dashed in blue) plasma components in the process of relaxation.

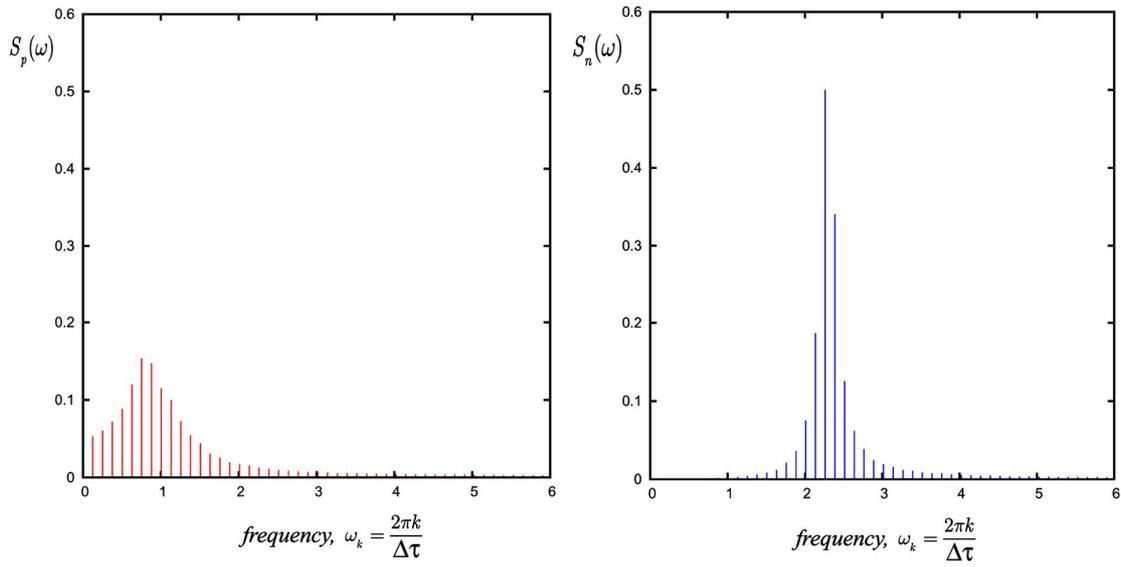

Fig.12. Spectral distributions of kinetic energy of drift motion for positive (left) and negative (right) plasma components for coherent vortex state evaluated over the time interval $\Delta\tau = 350\ldots400$.

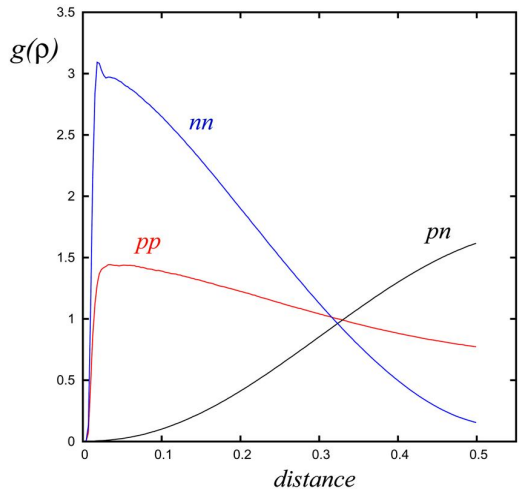

Fig.13. Radial distributions for coherent vortex state formed after the period of relaxation averaged over the time span $\tau = 350\ldots400$.